\documentclass[aip,apl,reprint]{revtex4-1}
\usepackage{amsmath}
\usepackage{amsfonts}
\usepackage{amssymb}
\usepackage{graphics}
\usepackage{float}
\usepackage{graphicx}
\usepackage{epstopdf}
\usepackage[compact]{titlesec}
\usepackage{natbib}
\usepackage{threeparttable}
\usepackage[titletoc,title]{appendix}
\usepackage{lipsum}
\usepackage{xcolor}
\begin{document}
\title{Photo-modulation of the spin Hall conductivity of mono-layer transition metal dichalcogenides}
\author{Parijat Sengupta}
\author{Enrico Bellotti}
\affiliation{Dept. of Electrical and Computer Engineering, Boston University, Boston, 02215.}

\begin{abstract}
We report on a possible optical tuning of the spin Hall conductivity
in mono-layer transition metal dichalcogenides. Light beams of
frequencies much higher than the energy scale of the system (the
\textit{off-resonant} condition) does not excite electrons but
rearranges the band structure. The rearrangement is quantitatively
established using the Floquet formalism. For such a system of
mono-layer transition metal dichalcogenides, the spin Hall
conductivity (calculated with the Kubo expression in presence of
disorder) exhibits a drop at higher frequencies and lower
intensities. Finally, we compare the spin Hall
conductivity of the higher spin-orbit coupled WSe$_{2} $ to MoS$_{2}
$; the spin Hall conductivity of WSe$_{2}$ was found to be larger.
\end{abstract}
\maketitle

The transition metal dichalcogenides (TMDCs) are layered materials
of covalently bonded atoms held together by weak van der Waals
forces~\cite{wang2012electronics} and represented by the formula
MX$_{2}$ where \textit{M} is a transition metal element from group
IV-VI while \textit{X} denotes the chalcogens S, Se, and Te. The
bulk TMDC when mechanically exfoliated gives a layered
two-dimensional configuration of atoms with distinct
characteristics.~\cite{huang2013metal} For instance, a single layer
of TMDC has high electron mobility, a direct band gap, absence of
dangling bonds and can be stacked in vertical
layers~\cite{lee2013flexible} to form hetero-junctions with clean
interfaces. The dispersion of a single layer of TMDC also supports a
rich variety of condensed matter phenomena, notably the coupling of
the valley and electron spin degree of
freedom~\cite{xiao2012coupled} without an external magnetic field.
The coupling of the spin and valley degree of freedom is most easily
observed at the time reversed pair of valley-edges \textit{K} and
\textit{K$^{'}$} and in their immediate vicinity. Interestingly,
this dispersion, as shown in Ref.~\onlinecite{tahir2014photoinduced}
can be modulated by light under off-resonant
conditions~\cite{lopez2015photoinduced} to give rise to a
valley-dependent tuning of the band gap and an overall alteration of
the carrier transport characteristics.

In this letter, we focus on spin currents in mono-layer TMDCs through
a quantitative evaluation of the inter-band spin Hall conductivity (SHC). Spin
currents can be generated in solid-state systems via spin-dependent
scattering from charged impurities due to spin-orbit (so) coupling-
the extrinsic spin Hall effect (SHE) or through a band
structure modification using built-in fields aided by so-interaction, commonly termed the intrinsic SHE. The SHE is a standard method to generate and detect spin currents and is
usually manipulated with an external electric field. We present an
alternative approach where an external light source modulates the
spin current (SHE-generated) which manifests as a change to the SHC. A sufficiently strong spin-orbit coupling is
however necessary to induce a tangible deflection of the carriers
based on their intrinsic spin polarization. The choice of mono-layer
TMDCs as a test bed for our work is driven by the fact that their
spin response properties exhibit an intermediate behavior between
the one observed for graphene with massless Dirac fermions and an
ordinary system of conventional 2D electron gas. In particular, the hole-doped
system shows markedly different spin response behavior while an
electron-doped mono-layer is closer to a 2D electron
gas.~\cite{hatami2014spin} We specifically choose MoS$_{2}$ and
WSe$_{2}$, since of all the known TMDCs that are semiconductors,
they have the lowest $\left(0.18\, eV\right)$ and highest spin-orbit
coupling $\left(0.46\, eV\right)$, respectively. This wide variance
in spin-orbit coupling may therefore help underscore its value in
the production of a pure spin-current via SHE.

Furthermore, while literature on optical control of quantum
transport in semiconductors exist~\cite{yao2007optical}, we show
here how light beams that operate in a distinct
\textit{off-resonant} state rearrange the energy dispersion and
modulate the overall spin-transport behavior. The
\textit{off-resonant} condition which is a non-equilibrium state happens when energy of the incoming radiation is much larger
than the hopping amplitude of the electrons ($ \hbar\omega \gg t $);
for such a case real photon absorption is inhibited by energy conservation,
instead second order virtual photon emission or absorption occurs
leading to reordering of electronic bands and photon-dressed
eigen-states.~\cite{lopez2015photoinduced,kitagawa2011transport} We theoretically calculate the SHC within the linear response framework for such a system and in presence of surface disorder to gauge the magnitude of the spin current. As we show later, the SHC declines as the frequency of the incoming radiation is raised or the intensity is lowered.

The basis for all calculations in this paper is the low-energy $ 4 \times 4 $ Hamiltonian shown in Eq.~\ref{mos2ham} for mono-layer TMDCs.
\begin{equation}
H_{\tau} = at\left(\tau\,k_{x}\sigma_{x} +  k_{y}\sigma_{y}\right)\otimes\mathbb{I} + \dfrac{\Delta}{2}\sigma_{z}\otimes\mathbb{I} - \dfrac{\lambda\tau}{2}\left(\sigma_{z} - 1\right)\otimes\,s_{z},
\label{mos2ham}
\end{equation}
The $ 4 \times 4 $ Hamiltonian describes two non-interacting $ 2
\times 2 $ blocks where the upper (lower) block furnishes the
dispersion of the spin-up (down) conduction and valence bands. The
index $ \tau = \pm\,1 $ distinguishes the two valley edges $
K\left(+\right) $ and $ K^{'}\left(-\right) $, $ a $ is the lattice
constant, and $ t $ denotes the hopping parameter. The energy gap
between the conduction and valence bands in absence of intrinsic
spin-orbit coupling is $ \Delta $. The Pauli matrices $
\hat{\sigma}_{i} $ where $ i = \left\lbrace x, y, z\right\rbrace $
act on the lattice sub-space while $ \hat{s}_{z} $ is linked to the
spin of the electrons at the non-equivalent high-symmetry points $ K
$ and $ K^{'} $ that are related through time reversal
symmetry.~\cite{zhu2014study,konabe2014valley}. The dispersion in
momentum space close to $ K $ and $ K^{'} $ edges can be written as
\begin{equation}
E_{\mu,\tau}^{\nu} = \dfrac{1}{2}\biggl[\mu\tau\lambda + \nu \sqrt{\left(\Delta - \mu\tau\lambda\right)^{2} + 4a^{2}t^{2}k^{2}}\biggr],
\label{eigval}
\end{equation}
where $  \mu = +\left(-\right) $ for the spin-up (down) polarization
and $ \nu = 1\left(-1\right) $ for the conduction (valence) band.
Note that the finite spin-orbit coupling, $ 2\lambda $, splits the
valence bands while the conduction states remain spin degenerate at
the edges $ K $ and $ K^{'} $. Using this Hamiltonian, we
proceed to examine the influence of off-resonant light and the
consequent distortion of the Bloch states including the band gap.

\begin{figure}
\includegraphics[scale=0.65]{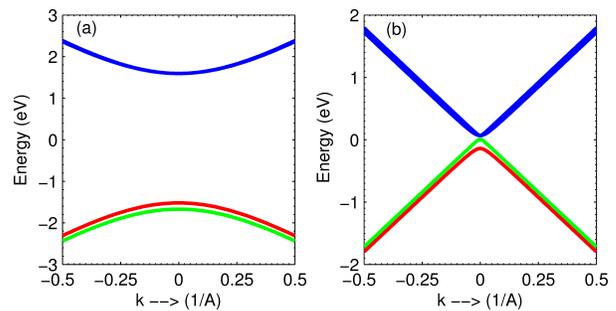}
\caption{The dispersion of mono-layer MoS$_{2}$ under off-resonant light condition. The sub-figure on the left (right) plots the band dispersion around the $ K (K') $ point.}
\label{disp_altered}
\end{figure}

The influence of the periodic off-resonant light on the TMDC
mono-layer is to the lowest order approximated by an effective
Hamiltonian averaged over a complete cycle through the evolution
operator $ U =
\mathcal{T}exp\left(-i\int_{0}^{T}H\left(t\right)dt\right)
$.~\cite{kitagawa2011transport} Here $ \mathcal{T} $ is the
time-ordering operator and $ T = 2\pi/\omega $. This approximate
Hamiltonian, which in principle describes the behaviour of a system
with time scales much longer than $ T $, rearranges the electron
occupation number without modifying the bands. Moreover, the system
under this approximation is transformed from the non-equilibrium
time-dependent case in to a static problem described by a stationary
effective Floquet
Hamiltonian.~\cite{kitagawa2011transport,cayssol2013floquet} The
Floquet Hamiltonian~\cite{tannor2007introduction} in general governs the
evolution of a quantum time-dependent periodic system through a
Schr{\"o}dinger equation which admits solutions of the form $
\vert\Psi_{\alpha}\left(t\right)\rangle =
exp\left(-i\varepsilon_{\alpha}t/\hbar\right)\vert\Psi_{\alpha}\left(t\right)\rangle
$, where the common periodicity of the system and the external
driving pulse is expressed as $
\vert\Psi_{\alpha}\left(t\right)\rangle = \vert\Psi_{\alpha}\left(t
+ T\right)\rangle $ and $ T = 2\pi/\omega $. For our purpose, in the
off-resonant state, the approximate Floquet Hamiltonian following
Ref.~\onlinecite{kitagawa2011transport} is
\begin{equation}
H_{\mathcal{F}} = H_{\tau} + \dfrac{1}{\hbar\,\omega}\left[H_{-1}, H_{1}\right],
\label{flham}
\end{equation}
and $ H_{m} =
\dfrac{1}{T}\int_{0}^{T}H\left(t\right)exp\left(-im\omega t\right)dt
$. Note that $ H\left(t\right) $ is the time-dependent part obtained
using the standard Peierl's substitution $ \hbar\,k\rightarrow
\hbar\,k - e\textbf{A}\left(t\right)$ in the TMDC mono-layer
Hamiltonian (Eq.~\ref{mos2ham}); this substitution gives $
H\left(t\right) =  \dfrac{at}{\hbar}\,A\left(\sigma_{x}cos\,\omega t
+ \sigma_{y}sin\,\omega t\right) $, where the off-resonant light is
right-circularly polarized and represented through the vector potential $
\textbf{A}\left(t\right) =  A\left(cos\,\omega t\,\hat{e}_{x}, sin\,\omega
t\,\hat{e}_{y}\right) $. The amplitude and frequency are denoted by $ A $ and $
\omega $, respectively. The desired Floquet Hamiltonian, $ H_{\mathcal{F}} $, by
a direct evaluation of the respective Fourier components and using $
\left[\sigma_{x}, \sigma_{y}\right] = 2i\sigma_{z} $ therefore reads
similar to Eq.~\ref{mos2ham} but with a different band gap. The
change in band gap by evaluating the commutator in
Eq.~\ref{flham} and inserting in Eq.~\ref{mos2ham} is
expressed as $ \left(\Delta/2\right)\sigma_{z}\otimes\mathbb{I}
\rightarrow \left[\left(\Delta +
\tau\,\Delta_{F}\right)/2\right]\sigma_{z}\otimes\mathbb{I} $, where $
\Delta_{F} = 2e^{2}A^{2}a^{2}t^{2}/\hbar^{3}\omega $ is the
light-induced band gap modification and $ A = E_{0}/\omega $ with $
E_{0} $ being the amplitude of the electric field. We clarify that all future references to the band gap, $ \Delta $, in the text from now shall includes the light-induced modulation.  Note that for brevity we have only shown the Floquet induced change to the upper $ 2 \times 2 $ block of the Hamiltonian in Eq.~\ref{mos2ham}. A more
convenient representation utilizing the relation $ at = \hbar\,v_{f}
$ allows us to write this as $ 2\left(eAv_{f}\right)^{2}/\hbar\omega
$. This light-induced band gap under off-resonant conditions is
evidently alterable through the intensity and frequency parameters. To
see this, recall that the intensity of incident light is $ I =
\left(eA\omega\right)^{2}/\left(8\pi\alpha\right)$, $ \alpha = 1/137
$ denoting the fine structure constant.~\cite{zhai2014photoinduced}
The Floquet modulated band gap can therefore be rewritten as $
16\pi\alpha\,Iv_{f}^{2}/\omega^{3} $. The dispersion diagram when
right-circularly polarized light (under \textit{off-resonant} conditions) shines on a mono-layer of MoS$_{2}$ with altered band gaps is shown in Fig.~\ref{disp_altered}. Notice
that the band gap at $ K $ is increased to $ 3.11\, eV $ from the
pristine $ 1.66\,eV $ while its time-reversed counterpart at $ K^{'}
$ sees a reduction to $ 0.074\,eV $ for right-circularly polarized
light. The enhancement and reduction at the valley edges is reversed
for a left-circularly polarized beam. The energy of the light
beam was assumed to be $ eAv_{f} = 2.9\, eV $.  This result is in qualitative agreement with Ref.~\onlinecite{tahir2014photoinduced}.

We have until now obtained the desired form of the Floquet
Hamiltonian for a mono-layer TMDC; for the next stage of
calculations involving SHC, the Fermi level is
positioned to the top of the valence bands such that the conduction
bands are devoid of carriers. The Kubo expression (for
non-interacting particles) for SHC is written
using the eigen states and function of the representative
Hamiltonian as
\begin{equation}
\sigma_{SH} = -i\dfrac{\hbar\,e}{L^{2}}\sum_{n,n^{'}}\dfrac{f\left(\varepsilon_{n}\right)- f\left(\varepsilon_{n^{'}}\right)}{\varepsilon_{n} - \varepsilon_{n^{'}}}\dfrac{\langle\,n\vert\,j_{x}^{z}\vert\,n^{'}\rangle \langle\,n^{'}\vert\,\hat{v}_{y}\vert\,n\rangle}{\varepsilon_{n} - \varepsilon_{n^{'}}+i\zeta},
\label{kubof}
\end{equation}
where we choose $ \vert\,n\rangle $ and $ \vert\,n^{'}\rangle $ to
be the valence and conduction eigen functions of the Hamiltonian
given in Eq.~\ref{flham}. We also write down the corresponding
energy difference for later use using Eq.~\ref{eigval} as $
\epsilon_{n^{'}} - \epsilon_{n} = \sqrt{\left(\Delta -
\lambda\right)^{2} + 4a^{2}t^{2}k^{2}} $. We explicitly state here
that the wave functions are for the $ K^{'} $ valley edge; the
choice of the $ K^{'} $ valley edge is dictated by the lower band
gap (compared to the $ K $ edge) which ensures a higher conductivity
response. At the $ K^{'} $ valley, the analytic representation of
the wave functions are $ \Psi_{VB} =
\dfrac{1}{\sqrt{2}}\begin{pmatrix}
\eta_{-}e^{-i\theta} \\
-\,\eta_{+}
\end{pmatrix} $ and $ \Psi_{CB} = \dfrac{1}{\sqrt{2}}\begin{pmatrix}
\eta_{+}e^{-i\theta} \\
\,\eta_{-}
\end{pmatrix} $ where $ \theta = \tan^{-1}k_{y}/k_{x} $ and $ \eta_{-} $ and $ \eta_{+} $ can be written as
\begin{equation}
\eta_{\pm} = \sqrt{1 \pm \dfrac{\Delta - \lambda}{\sqrt{\left(\Delta - \lambda\right)^{2} + \left(2atk\right)^{2}}}}.
\label{vbcoeff}
\end{equation}

\begin{figure}
\includegraphics[scale=0.45]{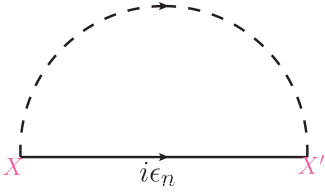}
\caption{The second order self energy term in the Born approximation after averaging over the impurity distribution. The labels $ X $ and $ X^{'} $ denote the placement of two interacting impurities while the dashed line represents their average.}
\label{feyn1}
\end{figure}

Since most TMDC mono-layers have impurities on the surface, their
role in adjusting the spin response must be accounted; in this work,
disorder is modeled as an effective retarded self-energy within the
self-consistent Born approximation (SCBA) that will allow us to
estimate the quasi-particle relaxation time and broadening of
states.~\cite{di2015statistical} The pair of SCBA equations
(for dilute disorder) being:
\begin{equation}
\begin{aligned}
\label{scba1}
G_{ks}\left(\epsilon\right) = \dfrac{1}{\epsilon - \epsilon_{ks} - \Sigma\left(\epsilon\right)};
\Sigma\left(\epsilon\right) = n_{i}v_{i}^{2}\int\,\dfrac{d^{2}k}{4\pi^{2}}G_{ks}\left(\epsilon\right),
\end{aligned}
\end{equation}
where $ n_{i} $ and $ v_{i} $ denote the density and strength of
impurities, respectively and $ G_{ks}\left(\epsilon\right) $ is the
retarded Green's function diagonal with respect to the band index
\textit{s} ($
\langle\,s\vert\,G_{k}\left(\epsilon\right)\vert\,s\rangle =
\delta_{ss^{'}}G_{ks}\left(\epsilon\right) $). The self-energy $
\Sigma $ which is also diagonal with respect to the band index
\textit{s} and independent of \textbf{\textit{k}} in SCBA is
averaged over impurity distributions and represented using a Feynman
diagram in Fig.~\ref{feyn1}. The single solid line in
Fig.~\ref{feyn1} is the unperturbed Green's function while the
combined dashed and solid line serves as the disorder-induced
effective self-energy in the lowest approximation.
The unperturbed retarded Green's function for the  $ 2 \times 2 $
upper block of the Hamiltonian in Eq.~\ref{mos2ham} is $ G_{0, R} =
\left(E - H_{\mathcal{F}}^{2 \times 2} + i\delta\right)^{-1} $ which when
expanded gives
\begin{equation}
G_{0,R} = \dfrac{1}{Det}\begin{pmatrix}
E + \dfrac{\Delta}{2}-\lambda + i\delta & -atk\exp\,-i\theta \\
-atk\exp\,i\theta & E - \dfrac{\Delta}{2} + i\delta
\end{pmatrix},
\label{retgreen}
\end{equation}
where $ Det $ is the determinant of the matrix. Inserting $ G_{0,R}
$ in the self-energy expression (Eq.~\ref{scba1}), we recast the
diagonal terms to the form $ \dfrac{1}{x \pm i0^{+}} $ to separate
the real and imaginary parts. The imaginary part of self-energy
supplies the inverse of the relaxation time while the real part
simply renormalizes the Fermi energy and is absorbed in the chemical
potential. Using the standard expression $ \dfrac{1}{x \pm i0^{+}} =
\mathbb{P}\dfrac{1}{x} \pm i\pi\delta\left(x\right) $, the
$ \delta\left(\cdot\right) $ functions under integration which lead to the
final self-energy imaginary terms are $ \delta\left(E + \Delta/2 -
\lambda - \left(atk\right)^{2}/\left(E - \Delta/2\right)\right) $
and $ \delta\left(E - \Delta/2 - \left(atk\right)^{2}/\left(E +
\Delta/2 - \lambda\right)\right) $. Neglecting the $ a^{2}t^{2}k^{2}
$ term in each of the $ \delta\left(\cdot\right) $ functions since
we must stay close to the valley edge ($ k $ is therefore a small
number and the product $ atk $ can be neglected), the imaginary part
of self-energy is
\begin{flalign}
Im\,\Sigma &= n_{i}v_{i}^{2}\int \dfrac{d^{2}k}{4\pi^{2}}\biggl[\delta\left(E + \Delta/2 - \lambda\right) + \delta\left(E - \Delta/2\right)\biggr], \notag \\
&\approx n_{i}v_{i}^{2}\dfrac{1}{2a^{2}t^{2}}\left(\vert\lambda - \Delta/2\vert\right).
\label{imself2}
\end{flalign}
Notice that $ \lambda - \Delta/2 $ happens to be the top of the
valence band which by assumption is also the position of the Fermi
level. The other argument of the $ \delta\left(\cdot\right) $ from
Eq.~\ref{imself2} is $ E = \Delta/2 $ which is the base of the
conduction band at the valley edges. This energy state is by
assumption above the set Fermi level and therefore discarded. Note
that the imaginary term in Eq.~\ref{kubof} is $ \zeta = \hbar/\tau_{tr} = 2Im\Sigma $. The carrier transit time is $\tau_{tr} $.

The spin Hall current that we calculate and has been addressed elsewhere~\cite{nomura2005nonvanishing} in context of a 2DEG with Rashba-coupling is essentially a non-equilibrium situation; an electric field applied along the $ \hat{y}-$axis gives rise to an out-of-plane ($ \hat{z}$-polarized) non-equilibrium $ \hat{x}$-directed spin current. The spin-current operator~\cite{murakami2006intrinsic} around $ K^{'}\left(\tau = -1\right) $ in this case is $ j_{x}^{z} = \dfrac{\hbar}{4}\left\lbrace \hat{v}_{x},\hat{s}_{z}\right\rbrace = -at\sigma_{x}s_{z}/2 $, where we have set the electron velocity operator along $ \hat{x}-$ and $ \hat{y-}$ axes as $\left(-at\hat{\sigma}_{x}/\hbar\right)$ and $\left(at\hat{\sigma}_{y}/\hbar\right)$, respectively. Note that the lattice space operators $ \hat{\sigma}_{x,y} $ commute with the Pauli spin operators $\left(\hat{s}_{x},\hat{s}_{y},\hat{s}_{z}\right)$. The matrix elements in Eq.~\ref{kubof}, utilizing the eigen states $ \vert\,n\rangle $ and $ \vert\,n^{'}\rangle $ therefore evaluate to
\begin{equation}
\begin{aligned}
\langle\,n\vert\,j_{x}^{z}\vert\,n^{'}\rangle &= \dfrac{at}{2}\left(\cos\theta - i\dfrac{\Delta - \lambda}{\epsilon_{n^{'}} - \epsilon_{n}}\sin\theta\right),  \\
\langle\,n^{'}\vert\,\hat{v}_{y}\vert\,n \rangle &= \dfrac{iat}{\hbar}\left(\cos\theta + i\dfrac{\Delta - \lambda}{\epsilon_{n^{'}} - \epsilon_{n}}\sin\theta\right).
\end{aligned}
\label{velmate}
\end{equation}
Notice that similar spin current operators for polarization and flow along a specific set of axes can also be written; for instance, $ j_{x}^{y} = \dfrac{\hbar}{4}\left\lbrace \hat{v}_{x},\hat{s}_{y}\right\rbrace = \hbar\tau\,at\sigma_{x}s_{y}/2 $.
\noindent By a direct substitution of the matrix elements from Eq.~\ref{velmate} followed by expanding the $ \sum $ in Eq.~\ref{kubof} (the TMDC mono-layer
sample area is assumed to be $ \mathcal{A} = L^{2} $) and
integrating out the angular part, we obtain the following expression
for the SHC in units of $ e/8\pi $, the universal SHC.~\cite{sinova2004universal}
\begin{equation}
\sigma_{xy}^{z} = 2a^{2}t^{2}\int_{0}^{k_{c}}\dfrac{\left(\Delta - \lambda\right)^{2} + \left(\epsilon_{n^{'}} - \epsilon_{n}\right)^{2}}{\left(\epsilon_{n^{'}} - \epsilon_{n}\right)^{2}\left[\left(\epsilon_{n^{'}} - \epsilon_{n}\right)^{2}+ \zeta^{2}\right]}k\,dk.
\label{shfinal}
\end{equation}
The SHC integral in Eq.~\ref{shfinal} can be numerically evaluated by choosing a cutoff radius in momentum space around the $ K^{'} $ high-symmetry edge..

\begin{figure}
\includegraphics[scale=0.8]{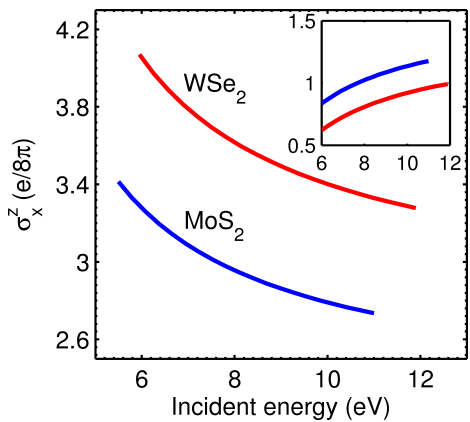}
\caption{The SHC of mono-layer MoS$_{2}$ and WSe$_{2}$ samples irradiated under off-resonant conditions for various incident energies at constant intensity. The higher spin-orbit coupling in WSe$_{2}$ results in an enhanced SHC. The inset shows the variation in band gap at the $ K^{'} $ edge of the mono-layer.}
\label{she1}
\end{figure}

For a numerical estimate of $ \sigma_{xy}^{z} $, we select band parameters~\cite{xiao2012coupled} for two mono-layer TMDCS, Mo$S_{2}$ and WSe$_{2}$. The incident
off-resonant illumination with frequencies much higher than the hopping
amplitude $ t $ must satisfy the
inequality~\cite{lopez2015photoinduced} $ \hbar\omega \gg t $; for
our case, we select a range of frequencies that lie within $ 5t \leq
\hbar\omega \leq 10t $. Additionally, to evaluate the retarded
self-energy, the impurity concentration was set to $ 2.5 \times
10^{10}\,cm^{-2} $ and the impurity potential was assigned the value of
$ 0.1\, keV\,\AA^{2} $.~\cite{adam2009theory} These numbers using
Eq.~\ref{imself2} yields a self-energy (imaginary contribution)
approximately equal to $ 4.6\, meV $ and $ 8.0\, meV $ for WSe$_{2}$
and MoS$_{2}$, respectively. Lastly, we choose the cutoff radius as $ k_{c} = 1.0\, 1/\AA $. The incident energy of
the beam ($ evA $) for all calculations has been held constant at $
2.0\,eV $ and as noted before is a tunable quantity via the
intensity $ I $. Inserting these numbers in Eq.~\ref{shfinal} and carrying out a numerical integration, the calculated $ \sigma_{xy}^{z} $ as a function of incident light energy is plotted in Fig.~\ref{she1}.
The SHC from Fig.~\ref{she1} is a fraction of
the universal spin Hall constant of $ e/8\pi $ and shows a declining
trend as the frequency increases which is easily explained by
resorting to the Floquet dependence of the band gap. The band gap
reduction for a constant energy beam is lower for a higher frequency
which essentially means a higher effective band gap at $ K^{'} $
edge. The inset in Fig.~\ref{she1} supports this reasoning. While we
have focused on band gap alterations through frequency modulation at
constant intensity, it is also possible to arrive at an identical
outcome by adopting the opposite. As a concrete example, the band
gap at the $ K^{'} $ edge under off-resonant conditions for two
equal energy {\bf ($ \hbar\omega = 10t = 11\, eV $)} light beams
whose incident energy is set to $ 2.0 \, eV $ and $ 3.0\, eV $, the
overall effective band gap is $ 0.854\, eV $ and $ 2.314\, eV $,
respectively. The corresponding SHC, as reasoned
for the case of higher frequency therefore drops with an increase
in the incident energy.

We want to point out that although the TMDC
mono-layer is considered pristine, edge defects and corrugations in
a real sample can quantitatively alter the final SHC through a
reduction in overall charge mobility (we ignore any possible
broadening of the local density of states). Further, the SHC is
theoretically temperature dependent, this dependence arises from the
change of mobility and carrier velocity. However, for device
operation and measurements performed in a small range of ambient
conditions (room temperature), we should expect a minimal shift
in mobility values for any discernible changes to spin
conductivity.

In conclusion, we have shown that the SHC in mono-layer TMDCs under
\textit{off-resonant conditions} is tunable via the intensity and
frequency of incident light. The procedure can be easily extended to
similar material systems that hosts massive Dirac fermions including
ultra-thin topological insulator films and other emerging 2D
materials such as black phosphorus.

This work was supported in part by the U.S. Army Research Laboratory
through the MSME Collaborative Research Alliance.

\bibliographystyle{apsrev}

\end{document}